\documentclass[a4paper]{article}
\usepackage[margin=2cm]{geometry}
\usepackage{url,lineno,microtype}
\usepackage[onehalfspacing]{setspace}

\usepackage{caption}
\usepackage{graphicx, amsmath,amssymb,amsfonts,booktabs}
\usepackage{textcomp,multirow,tabularx}
\usepackage[usenames,dvipsnames]{xcolor}
\graphicspath{{_figures/}}
\usepackage{placeins} 
\usepackage{hyperref}
\usepackage{footnote}
\hypersetup{colorlinks=true}
\usepackage{makecell}

\usepackage{array}
\newcolumntype{C}[1]{>{\centering\let\newline\\\arraybackslash\hspace{0pt}}m{#1}}
\newcolumntype{L}[1]{>{\raggedright\let\newline\\\arraybackslash\hspace{0pt}}m{#1}}
\newcolumntype{R}[1]{>{\raggedleft\let\newline\\\arraybackslash\hspace{0pt}}m{#1}}

\begin{document}

\title{\textbf{Using Shallow Neural Networks with Functional Connectivity from EEG signals for Early Diagnosis of Alzheimer's and Frontotemporal Dementia}}
\author{Zaineb Ajra$^{1}$, Binbin Xu$^{2}$, Gérard Dray$^{2}$, Jacky Montmain$^{2}$, Stéphane Perrey$^{1}$}
\date{\small $^{1}$EuroMov Digital Health in Motion, Univ. Montpellier, IMT Mines Ales, Montpellier, France. \\
$^{2}$EuroMov Digital Health in Motion, Univ. Montpellier, IMT Mines Ales, Ales, France.\\
\texttt{zaineb.ajra@umontpellier.fr, binbin.xu@mines-ales.fr}}
\maketitle

\begin{abstract}
  \textbf{Introduction: } Dementia is a neurological disorder associated with aging that can cause a loss of cognitive functions, impacting daily life. Alzheimer's disease (AD) is the most common cause of dementia, accounting for 50--70\% of cases, while frontotemporal dementia (FTD) affects social skills and personality. Electroencephalography (EEG) provides an effective tool to study the effects of AD on the brain. 
\textbf{Methods: } In this study, we propose to use shallow neural networks applied to two sets of features: spectral-temporal and functional connectivity using four methods. 
We compare three supervised machine learning techniques to the CNN models to classify EEG signals of AD / FTD and control cases. We also evaluate different measures of functional connectivity from common EEG frequency bands considering multiple thresholds. 
\textbf{Results and Discussion: } Results showed that the shallow CNN-based models achieved the highest accuracy of 94.54\% with AEC in test dataset when considering all connections, outperforming conventional methods and providing potentially an additional early dementia diagnosis tool.  \\
\url{https://doi.org/10.3389%2Ffneur.2023.1270405}
\end{abstract}

\section{Introduction}
Dementia is a neurodegenerative disease that results in the destruction of nerve cells in the brain, leading to various symptoms that affect cognition, emotion and movement \cite{oliveira2015nonpharmacological}. It becomes more common as people age, and can have many different effects, some of which may be reversible. 
Early-onset dementia, which occurs before the age of 65 years, is most frequently caused by Alzheimer's disease (AD) or frontotemporal dementia (FTD). In contrast, late-onset dementia occurs after the age of 65 \cite{lattante2015defining}. 
AD is characterized by amnesia, fluent aphasia, and visuospatial difficulties. On the other hand, FTD is characterized by changes in personality and behavior \cite{mendez1998behavioral}.
AD affects neurons and disrupts neurotransmitters responsible for storing memories and transmitting messages, while FTD causes degeneration in the frontal and anterior temporal lobes. However, neuropsychological tests aiming to differentiate between FTD and AD often yield uncertain or contradictory results \cite{perry2000differentiating}. It is crucial to distinguish between the two, as they impact different cortical regions and exhibit distinct clinical findings. 

Neuroimaging techniques have made a significant contribution to the identification of AD and FTD \cite{mantzavinos2017biomarkers}.
Over the past two decades, electroencephalography (EEG) has gained significant interest in clinical practice and research as a non-invasive tool for diagnosing dementia and determining its severity. EEG signals record the electrical activity of the brain, and have the potential to serve as a biomarker for AD and other neurodegenerative diseases \cite{biagetti2021classification}. 
EEG analysis involves extracting useful information from EEG signals using various features such as time, frequency, and time-frequency. 
The time domain analysis involves statistical features such as mean, median, and standard deviation. The frequency domain features involve decomposing the signal into different frequency sub-bands such as delta, theta, alpha, beta, and Low-gamma. These features are commonly used in machine learning algorithms for AD classification.
Time-frequency features involve using a spectrogram image as an alternative method for representing the characteristics of raw EEG data. The spectrogram displays the variation in energy values and frequency responses over time, using different magnitudes. In comparison to other manual or supervised feature extraction techniques, spectrograms are more effective in classifying signal/time-series data since they include more unknown and valuable features \cite{ramos2020feature}.

Functional connectivity (FC)  \cite{friston2011functional} is among the most commonly used techniques for studying brain function. It aims to characterize the observational similarity between different brain regions and how such similarity changes due to patients' pathology or even under different mental tasks. This kind of analysis has been applied in various experimental contexts, ranging from high-resolution fMRI data to more recently, EEG data, to provide more detailed temporal information. There is mounting evidence suggesting that Alzheimer's disease and various psychiatric disorders are associated with disruptions or enhancements in FC \cite{bullmore2009complex}.

The aim of this work is to improve diagnostic accuracy of dementia by exploring EEG signals using shallow Convolutional Neural Network (CNN) to classify subjects with AD, FTD, and healthy control (HC) group. Our approach involves feature extraction in both the time-frequency domain and functional connectivity analysis. We also compared our results with conventional classification methods, such as Linear discriminant Analysis (LDA), Support Vector Machine (SVM), and K-nearest neighbors (kNN), which rely on temporal and frequency feature extraction.

\section{ Data and models }

\subsection{Dataset}
In this study, we explored the publicly available dataset collected by Miltiadous et al. in \cite{miltiadous2021alzheimer}.
The original dataset consisted of 19 EEG channels, with a sampling rate of 500 Hz, recorded from the scalp of 88 participants, including 36 AD patients, 23 FTD patients, and 29 healthy control subjects. {The mean age and standard deviation (SD) for the AD group were 66.4 (SD = 7.9), for the FTD group were 63.6 (SD = 8.2), and for the HC group were 67.9 (SD = 5.4)}.
The cognitive decline and functional performance of patients with AD were evaluated using the Mini Mental State Examination (MMSE) score. The EEG signals were obtained from participants who were seated, relaxed, and had their eyes closed, following a clinical protocol. 
{The released data were subjected to initial pre-processing steps, including band-pass filtering within the frequency range of 0.5 to 45 Hz and the signals were re-referenced to A1-A2. Then, Artifact Subspace Reconstruction (ASR) was applied to remove bad data periods exceeding the maximum acceptable standard deviation of 17 for the 0.5 second window. To further investigate signal enhancement, Independent Component Analysis (ICA) was applied to the EEG signals, resulting in 19 distinct ICA components. As part of this process, elements classified as 'eye artifacts' or 'jaw artifacts' were automatically identified and removed.}
The average recordings duration for AD and FTD groups is approximately 13 minutes, ranging from 11 to 17 minutes. The recordings for the HC subjects lasted for an average of 13.8 minutes, ranging from 12.5 to 16.5 minutes.

\begin{figure}[!htb]
\centering
\includegraphics[width=0.8\textwidth]{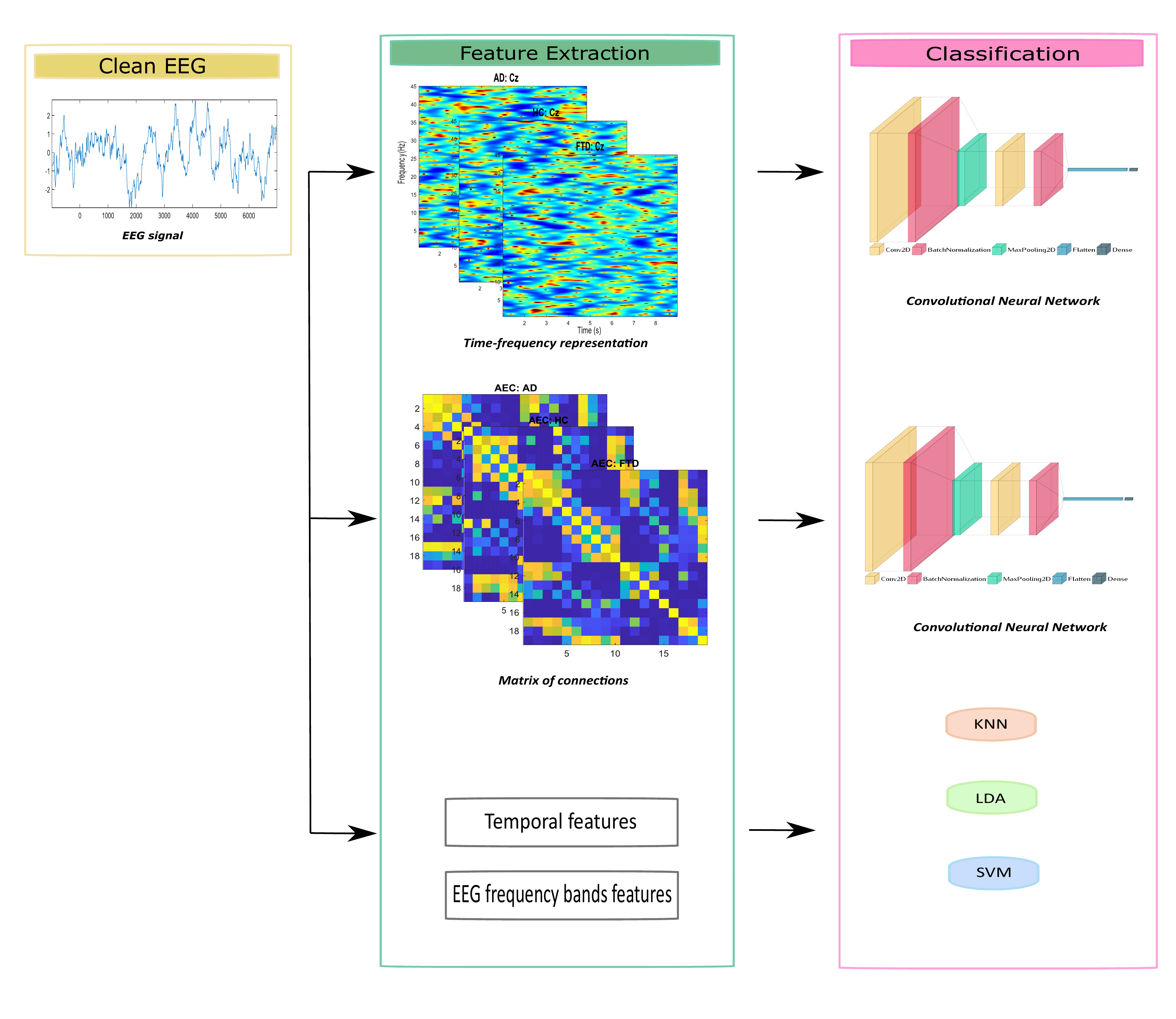}
\caption{Main workflow of this study.}
\label{fig:worflow}
\end{figure}

The main question of this work is how to improve the accuracy of dementia diagnosis by investigating EEG signals, so that we could achieve better classification rate on three groups: AD, FTD, and HC. We focus on feature extraction through functional connectivity methods using various thresholding techniques. Additionally, the study compares the outcomes with CNN based time-frequency analysis and traditional classification approaches as benchmarking against state-of-the-art methods. The overall workflow of this study is illustrated in Figure \ref{fig:worflow}. 
In the following section, we will detail the feature extraction steps and the classification methods. 

\subsection{Feature Extraction Methods}

\subsubsection{Spectral-temporal feature extraction}

\begin{figure}[!ht]
\centering
\includegraphics[width=0.8\textwidth]{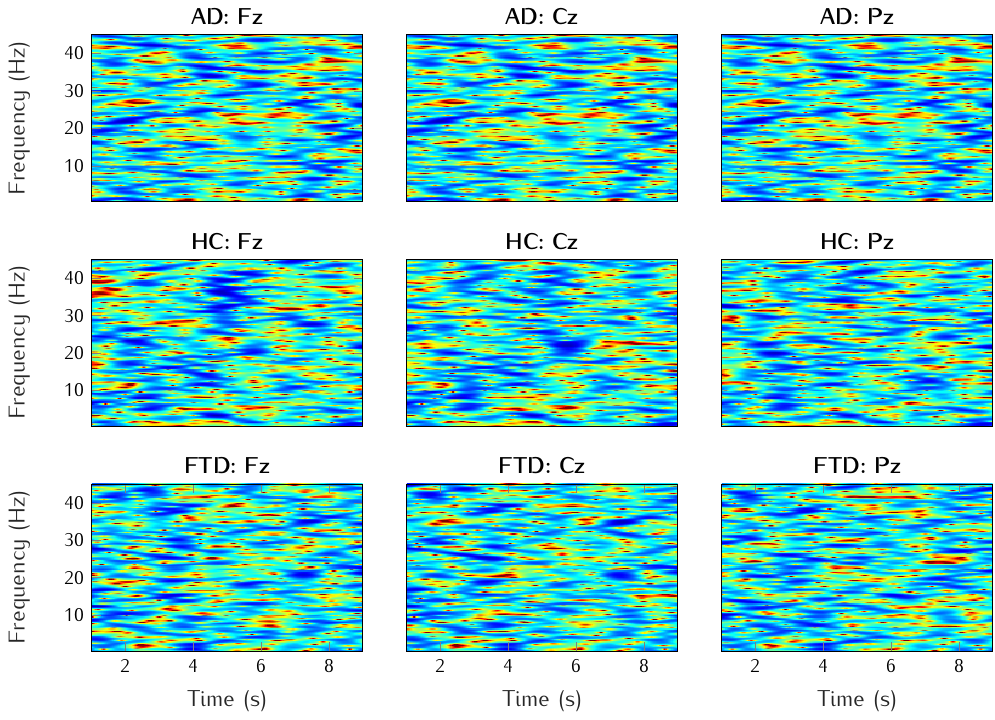}
\caption{Examples of the three classes (AD, FTD, HC), and three electrode channels (Fz, Cz, Pz)}
\label{fig:spectre}
\end{figure}

The time-frequency analysis allows to obtain richer information which are more appropriate to neural network \cite{Ajra2022Mental}. 
In the work of \cite{miltiadous2021alzheimer}, the EEG data was at first divided into epochs of 5 seconds with a 2.5 second overlap, features in the time and frequency domains were extracted for the classification.
In this study, we use different spectral-temporal feature extraction method: the preprocessed data was epoched of duration 10 seconds, then spectral-temporal features are extracted using Fourier Transform and Hanning window tapering approach with EEGLab (see examples in Figure \ref{fig:spectre}). 
In total, $130\,150$ spectrograms (subjects $\times$  channels $\times$ epochs) were extracted from the cleaned dataset. {The choice of the spectral-temporal features is supported by their ability to capture the temporal and spectral information of EEG signals. This approach is consistent with the existing literature, where similar methods have demonstrated effectiveness in classifying various forms of dementia. In \cite{fiscon2018combining}, the authors used a time-frequency features by applying both Fourier and Wavelet transforms to classify patients with Alzheimer’s disease versus healthy controls. This specific study achieved $83 \%$ accuracy using decision trees, highlighting the potential of time-frequency features in dementia classification.}
\subsubsection{Connectivity feature extraction}
\paragraph{Functional Connectivity Measures}
Functional connectivity refers to the interactions between brain regions, which can be quantified using measures of dependency between their temporal dynamics. By thresholding the connectivity values between all pairs of brain regions, one can identify functional connectivity networks that show which brain regions interact with each other. In EEG-based functional connectivity networks, nodes are represented by EEG channels, and links are connections between channel pairs \cite{bullmore2009complex}. The relationship between channels can be quantified using various methods, such as phase synchronization index (PSI), imaginary part of coherency (ImCoh), Pearson correlation (Corr) and Amplitude Envelope Correlation (AEC) \cite{bruns2000amplitude} \cite{sun2019graph}.

{Functional connectivity features offer an effective way to extract relevant information from EEG signals, as demonstrated by previous studies \cite{adebisi2020classification} and \cite{dottori2017towards} which used functional connectivity methods to discriminate between different groups, including AD/HC, AD/FTD, and FTD/HC, using SVM models. These studies provided valuable insights, with accuracies ranging from $72.2 \%$ to $87.67 \%$. The use of functional connectivity features is motivated by their ability to unveil complex interactions within EEG data, enabling the identification of distinct patterns that can differentiate between various neurological conditions.}
This study employed four functional connectivity estimation methods, namely PSI, ImCoh, Corr and AEC to compare the performances of time- and frequency-domain measures in classifying AD, FTD and HC. 
For each participant, we generated a connectivity matrix between all pairs of electrodes using the four functional connectivity measures separately. We applied three thresholding strategies on each connectivity matrix:
\begin{enumerate}
\item an absolute threshold with one commonly fixed value of 0.7, 
\item a proportional threshold varied from $10 \%$ to $90 \%$ with steps of $10 \%$, and 
\item no thresholding (raw connectivity matrix without any thresholding). These resulting matrices were then compared to evaluate the effects of thresholding on the classification performance.
\end{enumerate}
We evaluated the effects of thresholding on classification performance by comparing the resulting matrices. 
A total of 6850 connectivity matrices were then generated for each functional connectivity method, and these matrices were used for classification using a shallow Convolutional Neural Network.

\subparagraph*{Phase Synchronization Index}
Phase Synchronization Index measures how two or more signals are synchronized in terms of their phase relationship. It ignores the effect of amplitude and detects the correlation between different signal pairs only takes only into account the instantaneous phase relationship between the signals. 
Supposing the instantaneous phases of two signals $x(t)$ and $y(t)$ being $\phi_x(t)$ and $\phi_y(t)$, then the phase synchronization index (PSI) is defined as:
\begin{equation}
PSI =  \frac{1}{T}\left|\sum_{t=1}^T \exp \{j(\phi_x(t)-\phi_y(t))\}\right|
\end{equation}
The PSI is sensitive to phase change and its value ranges from 0 to 1. The PSI value of 1 indicates strict phase locking between the signals. On the other hand, a value of 0 indicates that the phases are uniformly distributed and there is no synchronization between the signals.

Examples of connectivity matrices with these four methods can be found in the following Figure \ref{fig:fc_psi_demo}.
\begin{figure}[!t]
\centering
\includegraphics[width=\textwidth]{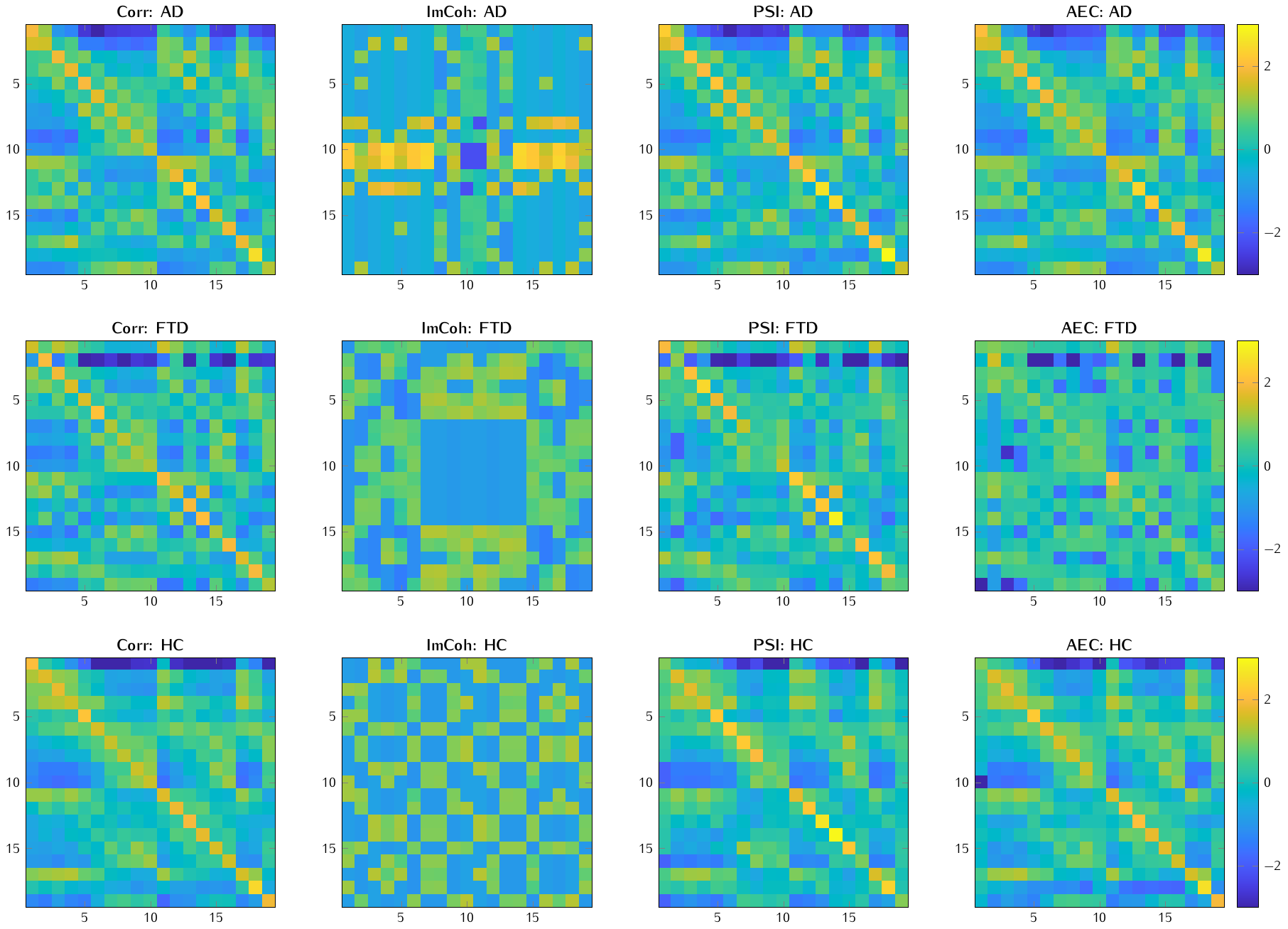}
\caption{Examples of connectivity matrices assessed by Corr, ImCoh, PSI and AEC for AD, FTD, and HC subjects, thresholded at 0.7 and \emph{z-score} normalized.}
\label{fig:fc_psi_demo}
\end{figure}

\subparagraph*{Imaginary Part of Coherency}
Coherency is a standard method to determine the spectral similarity between two signals. The coherency can be split into real and imaginary parts and different quantities can be analyzed further. Volume conduction in EEG recordings significantly affects coherence estimators. Electrical activity of the cortex disparately spreads across scalp electrodes at some distance from its generators allowing the same cortical activity to be measured by multiple neighboring electrodes at the same time i.e., with zero phases \cite{sakellariou2016connectivity}. 
This method is calculated from the coherency measure and is defined by : 
\begin{equation}
ImCoh = \frac{\left| I_{\text{mag}}(P_{xy}) \right|}{\sqrt{\left| P_{xx} \right| \left| P_{yy} \right|}}
\end{equation}

where $I_{\text{mag}}$ is the imaginary part of the power spectral density between the x and y signals, $P_{xy}$ and $P_{yy}$ denote the power spectral densities of x and y with themselves. This measure is between 0 and 1. If the value is close to 1, it means that there is a real link between the two signals. If the value is close to 0, it indicates that the signals are independent. Indeed, this link is realized by the volume conduction and therefore a false positive.

\subparagraph*{Pearson correlation coefficient}
Correlation (Corr) is used to estimate the level of linear dependence between two signals $x(t)$ and $y(t)$ in the time domain. The correlation is given by the following:
\begin{equation}
Corr =  \frac{Cov(x,y)}{\sigma_x \sigma_y}
\end{equation}
where $Cov(x,y)$ is the covariance between electrodes $x$ and $y$; $\sigma_x$ and $\sigma_y$ are the standard deviations of $x$ and $y$, respectively. Corr value varies between 1 and $-1$, where 1 is a complete positive linear correlation, 0 is no linear correlation, and $-1$ is a total negative linear correlation. The greater the absolute value of Corr becomes, the stronger the correlation is.

\subparagraph*{Amplitude Envelope Correlation}
The Amplitude Envelope Correlation (AEC) \cite{hipp2012large} is a commonly used method to measure the synchrony of cortical oscillations. This method involves calculating the Amplitude Envelope (AE) of a given cortical oscillation, which represents the energy fluctuations of the oscillation over time and is defined as the absolute value of the Hilbert transform. The AEC is then computed by correlating the AE of two oscillatory brain signals. High AEC values indicate synchronous AE fluctuations between oscillations or networks. Therefore, the AEC can provide important insights into the functional connectivity of different brain regions.

\paragraph{Graph Analysis} 

The core idea of functional connectivity can actually be considered as a network. One type of the most common tools to evaluate network is based on graph theory. 
Graph theory utilizes a mathematical framework to represent the connections between objects, where the objects are referred to as vertices, and the links that connect them are known as edges. 
The application of Graph theory to brain imaging data has demonstrated its potential as an understandable and adaptable method for representing brain networks \cite{bullmore2009complex}. 
Regarding brain networks in sensor space, the vertices in a graph can symbolize electrodes, and the edges can represent a certain measure of connectivity between these electrodes. The most common metrics to evaluate a graph are: mean degree (measures how interconnected the neighbors of a node are), clustering coefficient (a measure of local connectivity), efficiency (measures how efficiency information is transmitted across the graph) and betweeness centrality (quantifies the importance of a node in facilitating communication between other nodes). Usually, before computing graph metrics, to eliminate the background noise or other perturbations, random threshold or other justified thresholding is selected, and edge weights falling below this threshold are then adjusted to zero (and removed from the final graph) \cite{abazid2021comparative}.

\subsubsection{Conventional feature extraction}

Although the EEG signal's complexity makes it challenging to achieve clinically acceptable classification performance using feature engineering alone, it is still worthwhile to benchmark the performance of conventional machine learning models, as their computation costs are lower than those of deep learning models.
In this study, we extracted the conventional time and frequency domain features from each epoch to setup the classification dataset. Five basic EEG frequency bands are considered: delta, theta, alpha, beta, and Low-gamma. For the frequency domain features, we calculated the energy of each frequency band, while for the time-domain feature, we computed the minimum, maximum, mean, median, variance, standard deviation, kurtosis, and skewness.
{We considered these features because they have demonstrated their relevance in the EEG-based dementia classification task and their potential to capture distinct patterns. Many studies have used temporal analysis and energy assessments of EEG rhythms. For instance,\cite{tuauƫan2023tms} used temporal features from EEG signals (including maximum, minimum, mean, skewness, and kurtosis) as well as changes in signal energy. The authors achieved $83.1 \%$ accuracy in binary classification (distinguishing Alzheimer's patients from healthy controls) using the random forest classifier. Furthermore, our selection of frequency-domain features is supported by previous studies demonstrating their potential as informative markers for Alzheimer's disease classification. In particular, \cite{miltiadous2021alzheimer}, \cite{lindau2003quantitative} and \cite{safi2021early} have used dominant EEG frequency bands in their classification frameworks. These approaches yielded remarkable accuracies in distinguishing Alzheimer's patients from healthy subjects or frontotemporal dementia patients reaching $93 \%$ for a binary classification.}

\subsection{Classification}

\subsubsection{Convolutional Neural Network based models} 
Convolutional neural network (CNN) models are powerful tools for automatic feature extraction and classification, outperforming conventional machine learning methods \cite{Ajra2022Mental}. The effectiveness of CNN model design depends on several factors, including the number of layers, configurations, and training requirements. 2D CNN models excel at extracting two-dimensional features from images, potentially leading to superior performance.
In this study, we aim to classify two types of input: 
\begin{enumerate}
\item two-dimensional spectral-temporal features of dimension $224 \times 224$ from EEG signals, which can be treated as images. This size is the common size of state of the art CNN based models.  
\item functional connectivity matrices of dimension $19 \times 19$ extracted from each FC (19 EEG electrodes).
\end{enumerate}
To achieve this, we propose a shallow CNN model consisting of two blocks of 2D convolutional layers (Conv2D) that extract the most relevant features from the inputs. Each Conv2D layer is followed by a batch normalization layer, which speeds up training and improves model convergence. Both Conv2D layers use a Rectified Linear Unit (ReLU) activation and are equipped with 50 filters with a kernel size of $ 5 \times 5$ to disrupt the network's linear structure and make it sparse. The two blocks of Conv2D layers are connected by a max pooling layer with a size of $ 2 \times 2$, which reduces the number of parameters in the network and improves computational efficiency.

\subsubsection{Multi-frequency bands functional connectivity classification}
{In our study, we utilized a weighted approach to construct functional connectivity matrices. This involved applying a thresholding technique to the generated connectivity matrices. Thresholding is a widely employed method in research literature to eliminate weaker connections, as they are more susceptible to experimental noise \cite{stam2012organization}.
Two common approaches for thresholding in functional connectivity analysis are the absolute threshold and the proportional threshold.}
{The absolute threshold approach involves choosing edges with connectivity values exceeding a specified threshold value. On the other hand, the proportional threshold approach involves selecting the strongest percentage of connections within each network.}

{To classify the three groups (AD, FTD, and HC), we initially calculated the functional connectivity within the three groups using four different methods. We employ a thresholding technique on the resulting connectivity matrices, retaining only the highest connectivity values. In the case of a proportional threshold, we explore nine thresholding values ranging from $10 \%$ to $90 \%$ in increment of $10 \%$. Subsequently, based on the CNN model classification results, we determine the optimal threshold value from the nine options. Using this optimal threshold value, we generate four functional connectivity matrices in the four frequency bands (Theta (4–8 Hz), alpha (8–12 Hz), beta (12–30 Hz) and Low-gamma (30-45 Hz)) for the functional connectivity methods that showed the best classification performance (AEC, Corr and PSI). We then compare the classification performance of the optimal proportional value with that of the absolute threshold and with no thresholding applied.}

\subsubsection{Reference methods with Conventional machine learning algorithms}

In order to further evaluate the performances of the studies shallow CNN models, we also utilized conventional machine learning techniques to conduct epoch-based classification of three groups: AD, FTD and healthy control group. We utilized both temporal and frequency features in our approach. To address three distinct classification problems, we tested three commonly used classification algorithms in EEG studies: Linear Discriminant Analysis (LDA), Support Vector Machines (SVM), and K-Nearest Neighbors (kNN). We used a 10-fold cross-validation testing method to evaluate the performance of each algorithm. These three methods server as the baseline methods.

\subsection{Training and evaluation}
In summary, three feature sets were tested to discriminate between AD, FTD patients, and HC subjects: (1) statistical features, (2) functional connectivity features, and (3) time-frequency representation. Each set of features was used independently as input for the CNN model, which was trained using the same configuration for all sets of features. We utilized the Stochastic Gradient Descent with Momentum (SGDM) optimizer, with a learning rate fixed at $0.001$. The maximum number of training epochs was set to 50, with a batch size of 128. The same early stopping rule was applied, with a validation patience of 20 and a validation frequency of every 8 iterations.
Keeping the original subject / trial ratios from the raw data, samples from all features sets were randomly split into {\sc training} ($70\%$), {\sc validation} ($15\%$) and {\sc test} ($15\%$). 
To prevent over-fitting, 10 random triple-sets were generated and used for the CNN models. 
The median values of four metrics (Accuracy, Specificity, Sensitivity and F1-score) on {\sc test} sets are reported from the 50 epochs-long training.

The four specific quantitative indexes of the model’s classification performance are as follows:
\begin{equation}
\begin{aligned}
\text{Accuracy}    &=  \frac{TP + TN}{TP + TN + FP + FN} \\
\text{Specificity} &=  \frac{TN}{TN + FP} \\
\text{Sensitivity} &=  \frac{TP}{FN + TP} \\
\text{F1-score}    &=  2 \times  \frac{\text{Precision}   \times  \text{Sensitivity}}{\text{Precision}  + \text{Sensitivity}}
\end{aligned}
\end{equation}
Where Precision is defined as:
\begin{equation}
\text{Precision} =  \frac{TP}{TP + FP}
\end{equation}

{In these equations, TP denotes the correct classifications of positive cases, TN denotes the correct classifications of negative cases, FP denotes the incorrect classifications of negative cases into class positive, and FN denotes the incorrect classifications of positive cases into class negative.}

\section{ Results } 

\subsection{Classification performances: absolute threshold }

Table \ref{tab:model_perf}  presents the classification performance of different methods, where an absolute threshold of 0.7 for FC was used, and median accuracy is reported.
The three conventional machine learning models, SVM, LDA, and kNN, were tested under the same training conditions, and the results showed poor performance, with a global median accuracy of only at $58.68\% $, $ 57.89 \%$, and $ 59.77 \%$, respectively. These values were significantly lower than the results obtained with the CNN model. 
The CNN-based models were able to achieve an accuracy of $ 83.14 \%$, $ 87.82 \%$, $ 73.29 \%$, $ 85.87 \%$ and $ 81.97 \%$ for Time-Frequency analysis, PSI, ImCoh , Corr and AEC methods respectively. However, it is important to note that different approaches may be necessary to determine the brain regions involved since there's no universal method for determining functional connectivity \cite{bastos2016tutorial}. In fact, identifying the optimal method to infer these connections between brain regions remains a challenge in the field of network neuroscience \cite{shandilya2011inferring}. Regarding functional connectivity, the PSI method provided the best performance in the CNN model.

\begin{table}[!h]
\small
\setlength{\tabcolsep}{5pt}
  \centering
  \caption{The classification performance of different methods with an absolute threshold of 0.7 for FC (median $\%$). TF denotes Time-Frequency. The bold values represent the best results obtained by the proposed CNN.} 
    \begin{tabular}{lcccc}
    \toprule
    \textbf{Methods } & \textbf{ACC} & \textbf{Sensitivity} & \textbf{Specificity} & \textbf{F1 score} \\
    \midrule
    LDA  & 57.89  & 35.93 & 68.03  & 35.91 \\ 
    SVM & 58.68 & 37.03 &  67.98 & 36.94 \\ 
    KNN & 59.77 & 35.41  & 67.77 & 33.93 \\
    \midrule
    CNN-TF &  83.14 & 82.41 & 91.28 & 82.66  \\
    \midrule
    CNN-PSI & \textbf{87.82} & \textbf{86.89} & \textbf{93.43}  & \textbf{87.69} \\ 
    CNN-ImCoh & 73.29 & 71.07 & 86.87 & 72.60 \\
    CNN-Corr & 85.87 & 87.56 & 92.24 & 86.44 \\
    CNN-AEC & 81.97 & 84.83 & 90.15 & 83.24 \\
    \bottomrule
    \end{tabular}%
  \label{tab:model_perf}%
\end{table}%

Further result details regarding Table \ref{tab:model_perf} can be found in Table \ref{tab:model_perf_Group}. As observed, the choice of the FC used to create the matrices of connections has a significant impact on the overall predictive performance. For the AD group, Corr, PSI, and AEC yielded the best performance in the classification task with a sensitivity of $86.62 \%$, $86.15 \%$, and $85.21 \%$, respectively. Conversely, in the case of HC subjects, Corr outperformed the other methods. Additionally, among all functional connectivity methods, the FTD group exhibited the lowest classification performance, with sensitivities ranging between $76.64\%$ and $66.39 \%$.
Given that there's no universal approach to deduce connections and achieve precise results, it is crucial to explore various methods to establish a dependable framework for automatically diagnosing dementia.
Our proposed method, which involves using a matrix of connections as input to a CNN, provides more accurate results than conventional classification methods that require a prior feature engineering step. Since EEG signals are highly complex in nature, conventional techniques cannot always guarantee satisfactory classification performance. The reasons behind the high performance of our CNN-based approach can be attributed to the synergy between CNN and functional connectivity features. Functional connectivity features, which incorporate the complex interactions between distinct brain regions, offer a comprehensive representation of brain dynamics. Conventional methods often struggle to take advantage of this lack of information, relying primarily on manual feature engineering that can ignore crucial patterns. In contrast, CNN's hierarchical architecture excels at detecting patterns that consistently align with the complexities inherent in functional connectivity features. 
In addition, the integration of time-frequency features into the CNN learning process enriches its ability to understand the temporal and spectral aspects of EEG data. Overall, the strength of the CNN model lies in its capacity to capture features from a variety of data representations, eliminating the need for explicit feature engineering, which is particularly beneficial for complex signals such as EEG. The CNN's flexibility and ability to adapt to different types of data underline its superiority over traditional methods in this complex context.

\begin{table}[!htb]
\small
  \centering 
  \caption{The performances by groups with different methods, threshold of 0.7 for FC (median $\%$)}
  \begin{tabular}{c|ccc|ccc|ccc}
    \toprule
    \multirow{2}{*}{\textbf{Models}} & \multicolumn{3}{c|}{\textbf{Sensitivity}} & \multicolumn{3}{c|}{\textbf{Specificity}} & \multicolumn{3}{c}{\textbf{F1 score}} \\
    \cmidrule(lr){2-4} \cmidrule(lr){5-7} \cmidrule(lr){8-10} 
     & AD & HC & FTD & AD & HC & FTD & AD & HC & FTD \\
    \midrule
    LDA & 29.06 & 41.68 & 36.06 & 74.66 & 63.23 & 66.16 & 27.90 & 42.95 & 36.89 \\
    \addlinespace
    SVM & 47.25 & 28.83 & 35.02 & 58.68 & 80.33 & 64.95 & 45.58 & 29.36 & 35.89\\
    \addlinespace
    KNN & 38.67 & 57.21 & 10.34 & 63.38 & 46.28 & 93.66 & 37.69 & 48.63 & 15.48 \\
    \midrule
    CNN-TF &  85.81 & 83.30 & 77.67 & 88.02 & 90.68 & 94.69 & 84.84 & 83.15 & 79.95  \\
    \midrule
    CNN-PSI & 86.15 & 86.52 & 76.64 & 87.33 & 92.39 & 95.27 & 84.35 & 86.15 & 79.45 \\
    \addlinespace
    CNN-ImCoh & 77.58 & 71.91  & 66.39 & 81.50 & 84.93 & 92.20 & 75.99 & 71.91 & 69.68\\
    \addlinespace
    CNN-Corr & 86.62 & 87.92 & 74.59 & 88.00 & 92.09 & 95.52 & 84.52 & 86.46 & 78.51\\
    \addlinespace
    CNN-AEC & 85.21 & 84.27 & 73.36 & 85.67 & 91.12 & 95.14 & 82.63 & 83.73 & 77.14 \\
    \bottomrule
  \end{tabular}
 \label{tab:model_perf_Group}
\end{table}


\subsection{Classification performances: proportional threshold }

In functional connectivity analysis, absolute thresholding involves selecting edges with connectivity values above a fixed threshold and setting other edges to 0. This can lead to different network densities across subjects \cite{abazid2021comparative}. However, since the absolute threshold is fixed across all subjects, it may not be suitable for datasets with varying connectivity strengths or individual differences. To address this issue and ensure more comparable networks across subjects, we adopted in this work a proportional threshold. Therefore, by using a proportional threshold, we can achieve more consistent results and better account for individual differences in connectivity strengths, leading to a more reliable and interpretable comparison of functional connectivity networks across subjects.

Table \ref{tab:model_perf_proport_thre} presents the classification performance of different methods using a proportional threshold of the top $ 20 \%$ of connection values. For this method, we applied thresholding to the obtained real-valued connectivity matrices by selecting only the highest $20 \%$ of connectivity values. This method was used to ensure that we provide equal amounts of information to all functional connectivity matrices. Indeed, absolute thresholding used previously is a method of selecting a fixed threshold value to determine which connections in a network should be considered significant. However, the threshold selection is generally far from being an automated procedure, especially when using different functional connectivity metrics. Using a fixed threshold value across all analyses may not be appropriate for all methods and could potentially bias the results. Proportional thresholding, on the other hand, is a flexible thresholding method that adjusts the threshold value based on the distribution of the data. In this method, a percentage of the highest values of connections is selected as the threshold. The highest classification performance is observed with the Corr method, achieving an accuracy of $92.59 \%$, as shown in this table. On the other hand, the ImCoh shows the lowest classification performance with an accuracy of only $77.19 \%$.

\begin{table}[!ht]
\setlength{\tabcolsep}{5pt}
\small
  \centering
  \caption{The performances of different FC using a proportional threshold of keeping the highest $ 20\%$ of values (median $\%$). The bold values represent the best results obtained by the proposed CNN.} 

    \begin{tabular}{lcccc}
    \toprule
    \textbf{Methods } & \textbf{ACC} & \textbf{Sensitivity} & \textbf{Specificity} & \textbf{F1 score} \\
    \midrule
    CNN-PSI & 87.23 & 89.04 & 93.28  & 88.30 \\ 
    CNN-ImCoh & 77.19 & 74.59 & 86.27 & 76.47 \\
    CNN-Corr & \textbf{92.59} & \textbf{94.38} & \textbf{95.22} & \textbf{92.82} \\
    CNN-AEC & 89.86 & 89.91 & 95.07 & 90.12 \\
    \bottomrule
    \end{tabular}%
    \label{tab:model_perf_proport_thre}%
\end{table}

\subsection{Classification performances: varying thresholds }

\begin{figure}[!htb]
 \centering
 \includegraphics[width=0.5\textwidth]{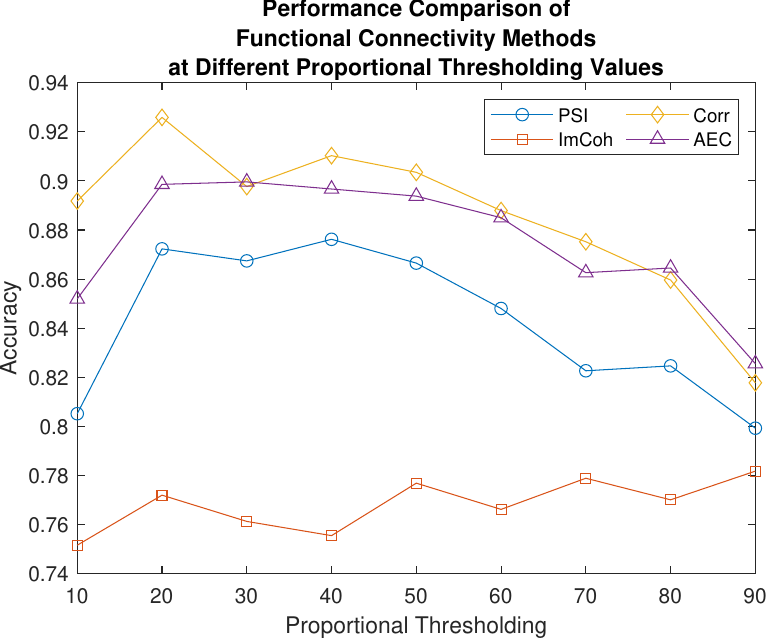}
\caption{Performance comparison of functional connectivity methods at different proportional thresholding values.}
\label{fig:FC_perf_curve_thre}%
\end{figure}

{In this section, we aimed to determine the most effective threshold for each connectivity method to enhance the classification accuracy of the CNN model, ultimately improving the accuracy and reliability of our functional connectivity-based classification approach. To achieve this, we applied a proportional thresholding on the obtained connectivity matrices from $10 \%$ to $90 \%$ with steps of $10 \%$. The performance of the CNN classifier in discriminating AD, FTD, and HC with the four functional connectivity methods is presented in Table \ref{tab:FC_perf_bythresholds}. 
Notably, the Corr and AEC methods demonstrated the highest classification performance, with accuracy values ranging from $82.55 \%$ to $92.59 \%$ across the nine proportional thresholding values.  
Particularly, we observe that retaining only $20 \%$ of the strongest connections yields the best classification performance, achieving an accuracy of $92.59 \%$ with Corr method. The comparison of the performance of functional connectivity methods across various proportional thresholding values is illustrated in Figure \ref{fig:FC_perf_curve_thre}.}

\begin{table}[htbp]
\small
  \centering
  \caption{Performance Evaluation of CNN Model with Proportional Thresholding (PT): Exploring 9 PT Values from 10\% to 90\% across Different Functional Connectivity Methods. The bold values represent the best results obtained by the proposed CNN.}
    \begin{tabular}{c | l | c | c | c | c}
    \toprule
    \textbf{Threshold} & \multicolumn{1}{c|}{\textbf{    Methods }} & \textbf{ ACC} & \textbf{ Sensitivity} & \textbf{ Specificity} & \textbf{ F1 score} \\
    \midrule
    \multirow{4}[0]{*}{10\%} & CNN-PSI  & 80.51 & 79.49 & 92.58 & 82.46 \\
          & CNN-ImCoh  & 75.15 & 72.75 & 87.46 & 74.11 \\
          & CNN-Corr  &  \textbf{89.18}  &  \textbf{90.17}  &  \textbf{94.76}  &  \textbf{90.53}  \\
          & CNN-AEC  & 85.19 & 86.52 & 93.73 & 86.91 \\
    \midrule
    \multirow{4}[0]{*}{20\%} & CNN-PSI  & 87.23 & 89.04 & 93.28 & 88.3 \\
          & CNN-ImCoh  & 77.19 & 74.59 & 86.27 & 76.47 \\
          & CNN-Corr  &  \textbf{92.59}  &  \textbf{94.38}  &  \textbf{95.22}  &  \textbf{92.82}  \\
          & CNN-AEC  & 89.86 & 89.91 & 95.07 & 90.12 \\
    \midrule
    \multirow{4}[0]{*}{30\%} & CNN-PSI  & 86.74 & 86.62 & 93.43 & 86.82 \\
          & CNN-ImCoh  & 76.12 & 76.97 & 85.67 & 74.36 \\
          & CNN-Corr  & 89.77 & 88.48 & 95.37 & 89.74 \\
          & CNN-AEC  &  \textbf{89.96}  &  \textbf{88.20}  &  \textbf{96.27}  &  \textbf{90.36}  \\
    \midrule
    \multirow{4}[0]{*}{40\%} & CNN-PSI  & 87.62 & 87.08 & 94.18 & 87.94 \\
          & CNN-ImCoh  & 75.54 & 73.77 & 87.01 & 74.84 \\
          & CNN-Corr  &  \textbf{91.03}  &  \textbf{92.42}  &  \textbf{95.82}  &  \textbf{91.73}  \\
          & CNN-AEC  & 89.67 & 88.97 & 84.17 & 90.24 \\
    \midrule
    \multirow{4}[0]{*}{50\%} & CNN-PSI  & 86.65 & 88.2  & 92.84 & 87.3 \\
          & CNN-ImCoh  & 77.68 & 75    & 86.57 & 75.63 \\
          & CNN-Corr  &  \textbf{90.35}  &  \textbf{88.20}  &  \textbf{96.93}  &  \textbf{90.89}  \\
          & CNN-AEC  & 89.38 & 89.89 & 94.63 & 89.89 \\
    \midrule
    \multirow{4}[0]{*}{60\%} & CNN-PSI  & 84.8  & 85.92 & 91.83 & 85.25 \\
          & CNN-ImCoh  & 76.61 & 76.97 & 86    & 74.36 \\
          & CNN-Corr  &  \textbf{88.79}  &  \textbf{88.76}  &  \textbf{93.88}  &  \textbf{88.64}  \\
          & CNN-AEC  & 88.5  & 87.64 & 95.37 & 89.27 \\
    \midrule
    \multirow{4}[0]{*}{70\%} & CNN-PSI  & 82.26 & 81.92 & 91.79 & 82.51 \\
          & CNN-ImCoh  & 77.88 & 73.88 & 89.85 & 76.56 \\
          & CNN-Corr  &  \textbf{87.52}  &  \textbf{87.56}  &  \textbf{93.73}  &   \textbf{88.18} \\
          & CNN-AEC  & 86.26 & 86.85 & 94.33 & 86.45 \\
    \midrule
    \multirow{4}[0]{*}{80\%} & CNN-PSI  & 82.46 & 80.9  & 92.97 & 83.43 \\
          & CNN-ImCoh  & 77    & 72.75 & 88.36 & 74.75 \\
          & CNN-Corr  & 85.96 & 87.09 & 92.69 & 86.89 \\
          & CNN-AEC  &  \textbf{86.45}  &  \textbf{87.08}  &  \textbf{93.88}  &  \textbf{87.02} \\
    \midrule
    \multirow{4}[0]{*}{90\%} & CNN-PSI  & 79.92 & 82.02 & 89.4  & 81.22 \\
          & CNN-ImCoh  & 78.17 & 76.12 & 87.16 & 77.16 \\
          & CNN-Corr  & 81.77 & 80.34 & 93.99 & 83.72 \\
          & CNN-AEC  &  \textbf{82.55}  &  \textbf{80.05}  &  \textbf{90.50}  &  \textbf{82.77} \\
    \bottomrule
    \end{tabular}%
  \label{tab:FC_perf_bythresholds}%
\end{table}%


\subsection{Classification performances without any thresholding on connectivity matrix}

Table \ref{tab:model_perf_weighted} reports the classification performance of different weighted FC without any thresholding. Interestingly, using raw (non-thresholded) connectivity matrices led to the highest classification performance, compared to using thresholded matrices. This means that setting a threshold may cause important information about connections between brain regions to be lost. Discarding weaker connections by setting a threshold can reduce the classifier's accuracy in distinguishing between groups. Additionally, the optimal threshold value can vary depending on the used FC method, making it difficult to choose an appropriate threshold. Therefore, using non-thresholded connectivity matrices may better represent the connectivity patterns and capture subtle differences between groups. We compared four functional connectivity methods and found that the AEC method was the most effective, while the ImCoh method was the least effective, in distinguishing between AD, FTD, and HC subjects.
\begin{table}[htbp]
\small
  \centering
  \caption{The performances of different weighted FC without any thresholding (median $\%$). The bold values represent the best results obtained by the proposed CNN.}
    \begin{tabular}{lcccc}
    \toprule
    \textbf{Methods } & \textbf{ACC} & \textbf{Sensitivity} & \textbf{Specificity} & \textbf{F1 score} \\
    \midrule
    CNN-PSI & 93.37 & 93.66 &  96.59 & 93.22 \\ 
    CNN-ImCoh & 77.10 & 78.09 & 85.50 & 75.85 \\
    CNN-Corr & 93.47 & 92.02 & 96.17 & 93.22 \\
    CNN-AEC & \textbf{94.54} & \textbf{95.22} & \textbf{96.72} & \textbf{94.56} \\
    \bottomrule
    \end{tabular}%
  \label{tab:model_perf_weighted}%
\end{table}%

\begin{figure}[!htb]
\centering
\includegraphics[width=0.9\textwidth]{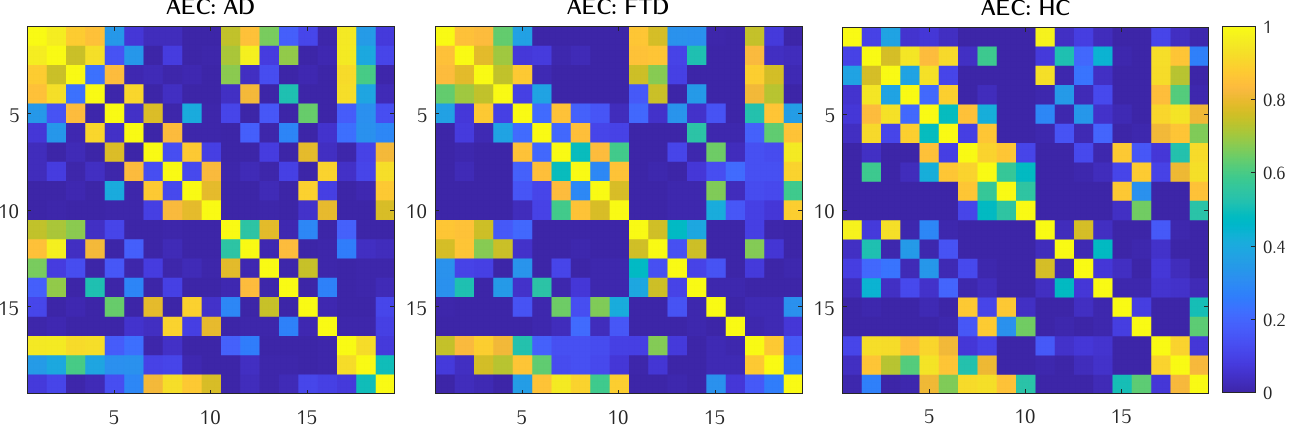}%
\\
\includegraphics[width=0.9\textwidth]{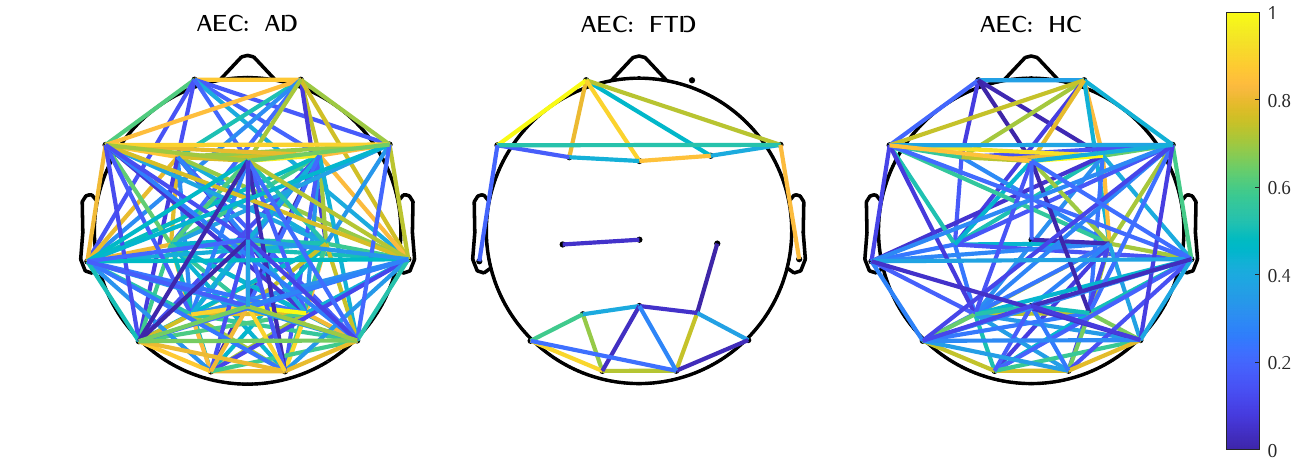}%
\caption{Mean FC matrices across all epochs (upper) and topologically significant local connectivity patterns among three groups: AD, FTD, and HC (lower)}
\label{fig:mean_fc_top_strength}
\end{figure}

When we compare FC matrices using AEC among AD, FTD, and HC subjects, we observed that FTD patients had stronger connections in frontal and temporal regions than HC subjects (Figure \ref{fig:mean_fc_top_strength}). These results align with a study that used fMRI to investigate functional connectivity in mild cognitive impairment (MCI) and AD patients compared to healthy controls  by \cite{penalba2022increased}. They found that AD patients had increased functional connectivity in the prefrontal cortex compared to healthy controls. Additionally, FTD patients showed disruptions in functional connectivity that were more widespread, particularly in regions affected by the disease, such as the frontal and temporal lobes \cite{hafkemeijer2017longitudinal}.

\subsection{Comparison of FC methods across various frequency bands}

{The division of EEG signals into frequency bands is a widely adopted approach in Alzheimer's disease research due to the distinct neural activity patterns observed at different frequency ranges. Certain frequency bands have been associated with specific cognitive functions and brain regions affected by the disease. For instance, abnormalities in the low-gamma frequency range have been linked to memory impairment and cognitive decline, while changes in delta and theta bands may indicate disruptions in attention and executive function. By examining EEG signals in frequency bands, several studies  \cite{abazid2022weighted, babiloni2020electrophysiology}, aim to capture these specific neural signatures related to Alzheimer's disease pathology, thus enhancing the ability of classification models to detect disease-related patterns.}

For the delta frequency band, the corresponding connectivity matrix has only extremely high values which make the classification fail. Therefore, we focus solely on the four other frequency bands (theta, alpha, beta, low-gamma) for analysis.

Tables \ref{tab:FC_FB_perf_nothreshold}--\ref{tab:FC_perf_FB_prop_thre} present the classification performance of the CNN model across different functional connectivity methods, considering various frequency bands and thresholding techniques.
Among the three threshold techniques, the CNN classification model utilizing functional connectivity measures without any thresholding demonstrated superior performance, surpassing the accuracy of the model employing functional connectivity methods with both absolute and proportional thresholds, achieving an accuracy of $60.33 \%$. 

\begin{table}[!htb]
\small
  \centering
  \caption{Performance comparison of FC methods across various frequency bands \emph{without any thresholding} (median $\%$). 
  The bold values represent the best results obtained by the proposed CNN.}
    \begin{tabular}{c | c | c | c | c | c}
    \toprule
    \textbf{FC method} & \textbf{Frequency band} & \textbf{ ACC } & \textbf{ Sensitivity} & \textbf{ Specificity} & \textbf{ F1 score} \\
    \midrule
    \multirow{4}[0]{*}{AEC}      & Theta  & 52.92 & 58.15 & 68.66 &  53.56 \\
          & Alpha  & 51.66 & 52.11 & 69.33 &  52.37 \\
          & Beta  & 54.09 & 49.53 & 62.61 &  55.31 \\
          & Low-gamma  &  \textbf{60.33}  &  \textbf{56.10}  &  \textbf{80.33}  &  \textbf{61.05}\\
    \midrule
    \multirow{4}[0]{*}{Corr}      & Theta  & 47.86 & 32.58 & 83.73 &  39.93 \\
          & Alpha  & 50.97 & 29.49 & 88.96 &  39.25 \\
          & Beta  & 51.36 & 51.4  & 72    &  49.39 \\
          & Low-gamma  &  \textbf{53.51}  &  \textbf{53.29}  &  \textbf{71.50}  &  \textbf{55.10} \\
    \midrule
    \multirow{4}[0]{*}{PSI}      & Theta  & 52.24 & 47.47 & 76.72 &  49.63 \\
          & Alpha  & 54.58 & 47.89 & 78.33 &  53.68 \\
          & Beta  & 57.12 & 52.53 & 80.6  &  55.57 \\
          & Low-gamma  &  \textbf{58.09}  &  \textbf{63.48}  &  \textbf{70.75}  &  \textbf{58.10} \\
    \bottomrule
    \end{tabular}%
  \label{tab:FC_FB_perf_nothreshold}%
\end{table}%

\begin{table}[!htb]
\small
  \centering
  \caption{Performance comparison of FC methods across various frequency bands \emph{ with an absolute threshold of 0.7} (median $\%$). The bold values represent the best results obtained by the proposed CNN.}
    \begin{tabular}{c | c | c | c | c | c}
    \toprule
    \textbf{FC method} & \textbf{Frequency band} & \textbf{ ACC } & \textbf{ Sensitivity} & \textbf{ Specificity} & \textbf{ F1 score} \\
    \midrule
    \multirow{4}[0]{*}{AEC}      & Theta  & 48.64 & 48.03 & 68.96 &  46.53 \\
          & Alpha  & 52.05 & 52.11 & 71    &  53.74 \\
          & Beta  & 50.1  & 47.19 & 71.34 &  46.93 \\
          & Low-gamma  &  \textbf{58.48}  &  \textbf{49.72}  &  \textbf{82.84}  &  \textbf{54.63} \\
    \midrule
    \multirow{4}[0]{*}{Corr}      & Theta  & 51.46 & 50.56 & 70.6  &  49.11 \\
          & Alpha  & 46.98 & 50.41 & 74.3  &  43.31 \\
          & Beta  & 46.1  & 37.36 & 75.52 &  40.74 \\
          & Low-gamma  &  \textbf{49.81}  &  \textbf{41.85}  &  \textbf{77.46}  &  \textbf{45.43} \\
    \midrule
    \multirow{4}[0]{*}{PSI}      & Theta  & 48.83 & 42.02 & 75.17 &  47.48 \\
          & Alpha  & 50.39 & 42.96 & 75.67 &  48.48 \\
          & Beta  & 50.39 & 39.75 & 88.49 &  45.01 \\
          & Low-gamma  &  \textbf{52.92}  &  \textbf{42.98}  &  \textbf{80.60}  &  \textbf{47.89} \\
    \bottomrule
    \end{tabular}%
  \label{tab:FC_perf_FB_absolute_thre}%
\end{table}%

\begin{table}[!htb]
\small
  \centering
  \caption{Performance comparison of FC methods across various frequency bands \emph{ using a proportional threshold of keeping the highest $ 20\%$ of values} (median $\%$). The bold values represent the best results obtained by the proposed CNN.}
    \begin{tabular}{c | c | c | c | c | c}
    \toprule
    \textbf{FC method} & \textbf{Frequency band} & \textbf{ ACC } & \textbf{ Sensitivity} & \textbf{ Specificity} & \textbf{ F1 score} \\
    \midrule
   \multirow{4}[0]{*}{AEC}       & Theta  & 44.74 & 30.28 & 82.33 & 39.09 \\
          & Alpha  & 42.11 & 38.2  & 65.67 & 37.67 \\
          & Beta  & 50.1  & 54.92 & 78.01 & 48.73 \\
          & Low-gamma  &  \textbf{56.82}  &  \textbf{41.80}  &  \textbf{86.42}  &  \textbf{50.62}  \\
    \midrule
    \multirow{4}[0]{*}{Corr}      & Theta  & 46    & 27.93 & 85.67 & 37.72 \\
          & Alpha  & 45.32 & 22.47 & 84.33 & 29.57 \\
          & Beta  & 48.25 & 53.05 & 67    & 48.37 \\
          & Low-gamma  &  \textbf{50.88}  &  \textbf{44.37}  &  \textbf{77.33}  &  \textbf{50.33}  \\
    \midrule
    \multirow{4}[0]{*}{PSI}      & Theta  & 45.61 & 29.81 & 84    & 39.14 \\
          & Alpha  & 36.55 & 36.52 & 71.94 & 38.58 \\
          & Beta  & 47.66 & 21.07 & 89.7  & 30 \\
          & Low-gamma  &  \textbf{48.44}  &  \textbf{50.84}  &  \textbf{67.76}  &  \textbf{48.07}  \\
    \bottomrule
    \end{tabular}%
  \label{tab:FC_perf_FB_prop_thre}%
\end{table}%

{In terms of classification performance of the three groups, the EEG Low-gamma band information exhibited the highest performance across all functional connectivity measures. Notably, when using the AEC method without any thresholding, a significant accuracy improvement was observed, resulting in an accuracy of $60.33 \%$. However, when comparing the classification of EEG frequency bands and the whole EEG spectrum, we can conclude that the classification of all EEG spectrum frequencies ranging from 0.5 to 45 Hz outperformed the classification based on individual EEG frequency bands. This suggests that considering the entire EEG spectrum provides valuable information for accurate discrimination of the three groups - AD, FTD, and HC - using the CNN model with functional connectivity measures.}

\subsection{Evaluation of metrics from Graph analysis }

\begin{figure}[!htbp]
\centering
\includegraphics[width=0.8\textwidth]{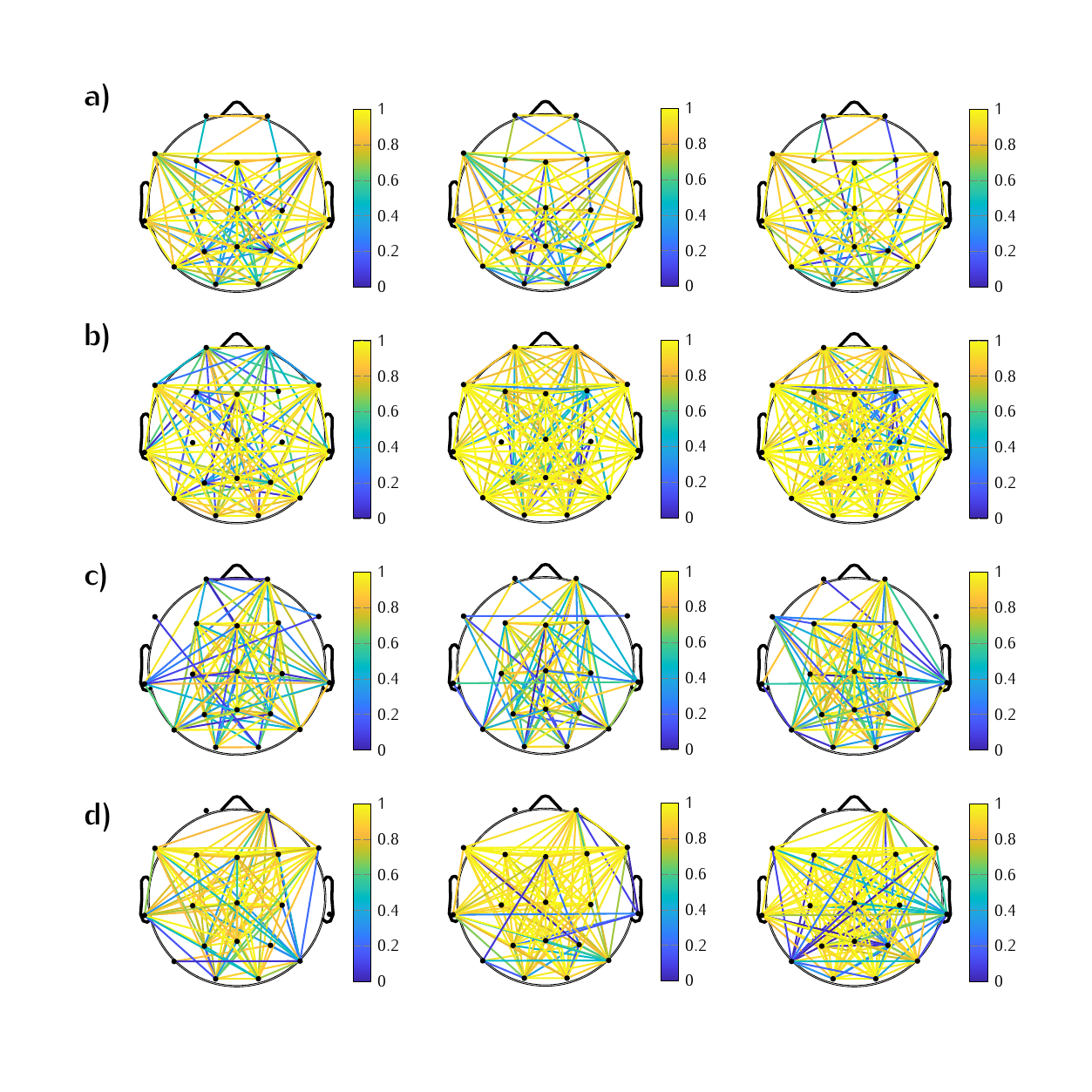}
\caption{Connectivity network of the AD (left), FTD (middle) and HC (right) groups on the a) \textit{theta}, b) \textit{alpha}, c) \textit{beta} and d) \textit{Low-gamma} bands with an absolute threshold of 0.7}
\label{fig:fc_aec_FB_absolute_thre}
\end{figure}

In addition to evaluating the performances of functional connectivity classification using the proposed CNN model, we have explored an alternative approach using the graph analysis, where graph metrics are employed to assess functional connectivity. This approach allows us to gain insights into the network-level properties and dynamics, providing a comprehensive understanding of the brain's functional organization in our study.
Here, we chose to present only the results of the Corr, PSI, and AEC methods for assessing graph properties in each EEG frequency band across the three groups. The decision to exclude the ImCoh method from the presentation of results is based on its lower classification performance when compared to the other three methods for all thresholding techniques and across all three groups.
In order to analyze the properties and characteristics of the brain network, Table \ref{tab:graph_metrics_AEC_absolute_thre} displays the graph metrics for functional connectivity networks in different frequency bands with an absolute threshold of 0.7 using AEC method for three distinct groups: AD patients, HC and FTD subjects. The graph metrics provide valuable insights into the topology and efficiency of brain networks in each group. Figure \ref{fig:fc_aec_FB_absolute_thre} illustrates the network connectivity between 19 electrodes of AD, FTD and HC groups on the Theta, Alpha, Beta and Low-gamma bands using AEC method and with an absolute threshold of 0.7. 

\begin{table}[!htb]
\setlength{\tabcolsep}{5pt}
  \small
  \centering 
  \caption{Comparison of Graph Metrics using AEC Method with an absolute threshold in AD, HC, and FTD Groups. The bold values represent the best results achieved for different graph metrics.
}
  \begin{tabular}{C{20mm}|ccc|ccc|ccc|ccc}
    \toprule
    \multirow{2}{20mm}{\textbf{Frequency bands}} & \multicolumn{3}{c|}{\textbf{Mean Degree}} & \multicolumn{3}{c|}{\textbf{Clustering Coefficient}} & 
    \multicolumn{3}{c|}{\textbf{Efficiency}} & \multicolumn{3}{c}{\textbf{Betweenness Centrality}} \\
    \cmidrule(lr){2-4} \cmidrule(lr){5-7} \cmidrule(lr){8-10} \cmidrule(lr){11-13}
     & AD & HC & FTD & AD & HC & FTD & AD & HC & FTD & AD & HC & FTD \\
    \midrule
    \addlinespace
    Theta & 11.89 & 10.95 & \textbf{17.05} & \textbf{0.93} & 0.89 & 0.91 & 0.65 & 0.62 & \textbf{0.83} & 13.05 & \textbf{13.15} & 2.94 \\
    \addlinespace
    Alpha & 13.68 & \textbf{15.89} & 12.84 & 0.92 & \textbf{0.99} & 0.98 & 0.68 & \textbf{0.80} & 0.68 & 5.36 & 5.26 & \textbf{10.73} \\
    \addlinespace
    Beta & 13.26  & 10.84 & \textbf{14.32} & 0.94 & 0.83 & \textbf{0.99} & 0.69 & 0.62 &  \textbf{0.73} & 9.57 & \textbf{12.10} & 8.10 \\
    \addlinespace
    Low-gamma & 13.05 & 12.42 & \textbf{13.47} & 0.91 & 0.88 & \textbf{1.00} & \textbf{0.64} & 0.61 & \textbf{0.64} & 3.89 & \textbf{5.15} & 2.94 \\
    \bottomrule
  \end{tabular}
 \label{tab:graph_metrics_AEC_absolute_thre}
\end{table}

Regarding the clustering coefficient metric, in the Alpha frequency band, the HC group shows the highest clustering coefficient, indicating a stronger tendency for nodes to form clusters in their brain networks. However, in the Beta and Low-gamma frequency bands, the FTD group have high clustering coefficients, indicating significant clustering in their brain networks.
Concerning efficiency metric, the FTD group exhibits the highest efficiency in the Theta and Beta frequency bands, implying that their brain networks facilitate information transmission more effectively compared to the AD and HC groups.
With reference to betweenness centrality metric, in Theta, Beta and Low-gamma frequency bands, the HC group consistently shows higher betweenness centrality than the AD and FTD groups, indicating a relatively higher influence of specific nodes in their brain networks.

{Overall, the results suggest that the FTD group tends to have more interconnected brain networks (higher mean degree) with efficient information transmission (higher efficiency) compared to the AD and HC groups. On the other hand, the HC group shows stronger clustering tendencies (higher clustering coefficient) and more prominent nodes facilitating communication between other nodes (higher betweenness centrality) in their brain networks.
These findings highlight the distinct characteristics of brain networks in each group, potentially providing valuable insights into the underlying neurodegenerative processes and the healthy brain's functional organization.}


\subsection{Exploring the Clinical Implications of Functional Connectivity Analysis}

In this paper, our analysis of functional connectivity holds significant promise for clinical applications in the early diagnosis and management of dementia. By examining brain network interactions via EEG signals, we can potentially identify subtle changes preceding cognitive decline, thus serving as sensitive biomarkers for early detection. The high accuracy achieved by our shallow neural network model in classifying AD/FTD/HC cases underlines the diagnostic potential of functional connectivity analysis. 

\begin{figure}[!htb]
\centering
\includegraphics[width=\textwidth]{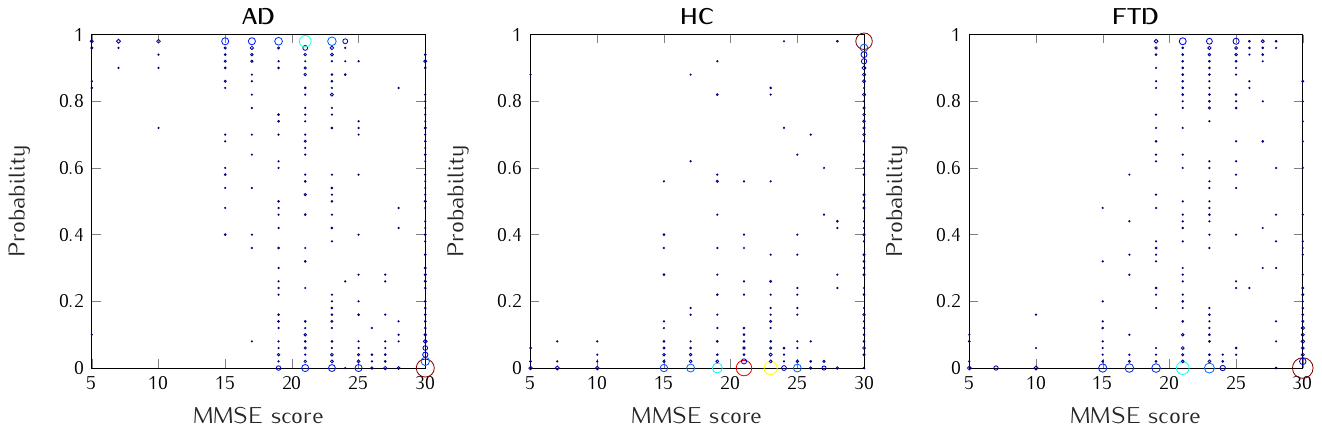}
\caption{Relationship between MMSE Scores and Prediction Probabilities for AD, HC, and FTD Groups, at trial level. All healthy subjects / patients included.}
\label{fig:fc_mmse_probability}
\end{figure}

In Figure \ref{fig:fc_mmse_probability}, we illustrate the relationship between the prediction probabilities of the AD, HC, and FTD groups and the MMSE scores identified in the testing set results.
In the AD group (left), a higher MMSE score was linked to a lower prediction probability, while among HC subjects (middle), a higher MMSE score was associated with an increased prediction probability. For the FTD group (right), a higher prediction probability was observed when the MMSE score ranged between 18 and 27. These findings align with clinical observations, where a lower MMSE score typically indicates a more pronounced cognitive decline. Notably, for the present dataset, the average MMSE score was 17.75 (SD = 4.5) for the AD group, 22.17 (SD = 8.22) for the FTD group, and 30 for the HC group.

Consequently, the classification probabilities assigned to each group could serve as a measurable indicator of dementia progression, providing clinicians with an objective index for monitoring disease progression and evaluating treatment. This information could not only enhance clinical decision-making, but also facilitate personalized treatment strategies and improve patient care in the field of dementia. Beyond diagnosis, these results hold promise for monitoring disease progression and evaluating treatment efficacy, offering clinicians with a versatile set of tools to improve early detection and patient-centered care in dementia disorders.

\subsection{Compared with Previous Studies}

{Table \ref{tab:Comparison of the classification accuracy} illustrates the comparison between the classification result in our study and that of previous EEG studies.
In recent years, several studies have explored the classification of dementia using EEG, employing various approaches. However, a particular study \cite{jiao2023neural}, stands out as it tested the performance of SVM and LDA models on a group of 890 subjects during their resting state. The study aimed to classify individuals into three levels: healthy controls (HC), mild cognitive impairment (MCI), and Alzheimer's disease (AD). The SVM model, utilizing features extracted from EEG such as absolute and relative power, Hajorth metrics, and time-frequency properties, achieved the highest accuracy of $70.2\%$. It's worth noting that most other studies focused on binary classification between different groups, such as AD/HC, FTD/HC, and AD/FTD. 
}

\begin{table}[htbp]
\small
  \centering
  \caption{Comparison of the classification accuracy of AD patients with other previous studies}
    \begin{tabular}[t]{ c | p{0.3\textwidth} | p{30mm} | p{20mm} | p{22mm}}
    \toprule
    \multicolumn{1}{c|}{\textbf{Study}} & \multicolumn{1}{c|}{\textbf{Feature set}} & \multicolumn{1}{c|}{\textbf{Classifier}} & \multicolumn{1}{c|}{\textbf{Classes}} & \multicolumn{1}{c}{\textbf{Best Acc}} \\
    \midrule
    2023   \cite{jiao2023neural} & Absolute power, relative power, Hjorth metrics (activity, mobility and complexity) and time-frequency property (STFT) & LDA, SVM & AD/MCI/HC & \makecell{{Acc = 70\%}, \\ {\tiny{(LDA)}} \\ Acc = 70.2\% \\ {\tiny{(SVM)}} } \\
    \midrule
    2023   \cite{tuauƫan2023tms} & Temporal features from EEG (maximum, minimum, mean, skewness and kurtosis), signal energy changes and TEP peaks & Random Forest & AD/HC & Acc = 83.1\% \\
    \midrule
    2023   \cite{prado2023source} & Functional connectivity features in source space from EEG & eXtreme Gradient Boosting & \makecell{AD/HC \\ AD/FTD} & \makecell{Acc = 87.1\% \\Acc = 86.7\%}\\
    \midrule
    2022   \cite{herzog2022genuine} & High order functional connectivity features from EEG source space & Random Forest & \makecell{FTD/HC \\ AD/HC} & \makecell{Acc = 93.15\% \\ Acc = 89\%}\\
    \midrule
    2022   \cite{miltiadous2021alzheimer} & Energy of EEG rythms, mean, variance and IQR features & Decision Trees, Random Forest, ANN, SVM, Naive Bayes and KNN   & \makecell{AD/HC \\ FTD/HC \\ AD/FTD}& \makecell{Acc = 99.1\% \\ Acc = 98\% \\ Acc = 91\%} \\
    \midrule
    2021   \cite{safi2021early} & Hjorth parameters EEG frequency bands, using DWT & SVM, RLDA and KNN  & AD/HC & \makecell{Acc = 97.64\% \\ {\tiny{(KNN)}}}\\
    \midrule
    2020   \cite{adebisi2020classification} & Functional connectivity features & SVM   & AD/HC & Acc = 87.67\% \\
    \midrule
    2018   \cite{fiscon2018combining} & Time-frequency features by applying both the Fourier and Wavelet transforms & Decision Trees & AD/HC & Acc = 83\% \\
    \midrule
    2017   \cite{dottori2017towards} & Functional connectivity features & SVM   & FTD/HC & Acc = 72.7\% \\
                 &       &       & AD/FTD & Acc = 72.2\% \\
                 &       &       & AD/HC & Acc = 44.9\% \\
    \midrule
    2011   \cite{nishida2011differences} & EEG rythms energy from source space & KNN   & \makecell{FTD/HC \\ {AD/FHC} \\ {FTD/AD} } & \makecell{Acc = 85.8\% \\ Acc = 92.8\% \\ Acc = 89.9\%} \\
    \midrule
    2003   \cite{lindau2003quantitative} & EEG frequency bands features (from delta and theta bands) & Logistic Regression & AD/FTD & Acc = 93.3\% \\
    \bottomrule
    \end{tabular}%
  \label{tab:Comparison of the classification accuracy}
\end{table}%
\FloatBarrier

\section{ Conclusion }

In this work, we propose an automatic diagnosis method for AD, FTD, and HC subjects using EEG time series and deep learning. We use four different approaches to infer the matrix of connections between brain areas: Phase Synchronization Index, Pearson's correlation, Imaginary part of Coherency, and Amplitude Envelope Correlation. These matrices are trained with CNN-based models.
In addition, we compared the performance of conventional methods, including SVM, LDA, and KNN, with the shallow CNN model for different feature extraction paradigms – time-frequency features and functional connectivity features. Our comparisons revealed that the CNN model outperformed conventional methods in dementia classification for both time-frequency and functional connectivity features.
The comparison of the four different approaches shows that our CNN-based method is more accurate, demonstrating the importance of network topology in describing brain data.
Our findings revealed that the CNN-AEC without any thresholding method is the most effective among the methods we studied, reaching $94.54 \%$ cross-validation accuracy.
The results suggest that EEG-based measures of functional connectivity, when combined with convolutional neural network, provide an accurate, reliable and rapid method of dementia classification and can significantly improve the efficiency of AD diagnosis.
The high performance of the basic CNN model suggests that a simple neural network architecture may be adequate for classifying dementia diseases. The pipeline is general and could be used for other mental disorder in which EEG time series can be recorded. 
In future work, it would be interesting to convert the EEG recordings captured at the scalp level into EEG time series data in the source space using source reconstruction. The CNN classifier will be assessed by comparing its performance in both the sensor space and the source space.

\section*{Acknowledgments}
ZA received a doctoral fellowship from AXIAUM Université Montpellier-ISDM (ANR-20-THIA-0005-01) and ED I2S in France.

\bibliographystyle{ieeetr}
\bibliography{bib}

\end{document}